\documentclass[fleqn, usenatbib]{mnras}

\usepackage{newtxtext,newtxmath}

\usepackage[T1]{fontenc}

\DeclareRobustCommand{\VAN}[3]{#2}
\let\VANthebibliography\thebibliography
\def\thebibliography{\DeclareRobustCommand{\VAN}[3]{##3}\VANthebibliography}

\usepackage{graphicx}	
\usepackage{amsmath}	

\title[No BH in NGC 2004 \#115]{NGC 2004 \#115: a black hole imposter containing three luminous stars}

\author[El-Badry, Burdge, \& Mr\'{o}z]{
Kareem El-Badry,$^{1,2,3}$\thanks{E-mail: kareem.el-badry@cfa.harvard.edu}  Kevin B. Burdge,$^{4,5}$
and Przemek Mr\'{o}z$^6$ \\
$^{1}$Center for Astrophysics $|$ Harvard \& Smithsonian, 60 Garden Street, Cambridge, MA 02138, USA\\
$^{2}$Harvard Society of Fellows, 78 Mount Auburn Street, Cambridge, MA 02138\\
$^{3}$Max-Planck Institute for Astronomy, K\"onigstuhl 17, D-69117 Heidelberg, Germany\\
$^{4}$Department of Physics, Massachusetts Institute of Technology, Cambridge, MA 02139, USA\\
$^{5}$Kavli Institute for Astrophysics and Space Research, Massachusetts Institute of Technology, Cambridge, MA 02139, USA \\
$^{6}$Astronomical Observatory, University of Warsaw, Al. Ujazdowskie 4, 00-478 Warszawa, Poland
}

\date{\vspace{-1.0cm}}

\pubyear{2021}

\begin{document}
\label{firstpage}
\pagerange{\pageref{firstpage}--\pageref{lastpage}}
\maketitle

\begin{abstract}
NGC 2004 \#115 is a recently identified black hole (BH) candidate in the Large Magellanic Cloud (LMC) containing a B star orbiting an unseen companion in a 2.9 day orbit and a Be star tertiary. We show that the unseen companion is not a 25\,$M_{\odot}$ BH, but a $(2-3)\,M_{\odot}$ luminous star. Analyzing the OGLE and MACHO light curves of the system, we detect ellipsoidal variability with amplitude 10 times larger than would be expected if the companion were a 25\,$M_{\odot}$ BH, ruling out the low inclination required for a massive companion. The light curve also shows a clear reflection effect that is well-modeled with a $2.5\,M_{\odot}$ main-sequence secondary, ruling out a lower-mass BH or neutron star companion. We consider and reject models in which the system is a binary containing a stripped star orbiting the Be star: only a triple model with an outer Be star can explain both the observed light curve and radial velocities. Our results imply that the B star, whose slow projected rotation velocity and presumed tidal synchronization were interpreted as evidence for a low inclination (and thus a high companion mass), is far from being tidally synchronized: despite being in a 2.9 day orbit that is fully or nearly circularized ($e < 0.04$), its surface rotation period appears to be at least 20 days. We offer cautionary notes on the interpretation of dormant BH candidates in binaries.
\end{abstract}

\begin{keywords}
stars: black holes -- binaries: spectroscopic -- stars: emission-line, Be
\vspace{-0.5cm}
\end{keywords}



\section{Introduction}
 
The search is on for detached stellar-mass BHs.
Recently, \citet{Lennon2021} reported the discovery of a $25\,M_{\odot}$ BH candidate in NGC 2004, a young (10-20 Myr) star cluster in the LMC. They observe a narrow-lined B star in a short-period ($P_{\rm orb}\approx 2.9$ days) orbit around a companion of uncertain nature. If the B star is interpreted as a main-sequence star, its inferred temperature and radius $(T_{\rm eff,B}\approx 22\,{\rm kK};\,\,R_{\rm B}\approx 5.6\,R_{\odot})$ imply a mass of about $8.6\,M_{\odot}$. The mass function of the orbit is unremarkable, $f_m\approx 0.07\,M_{\odot}$.  For a B star mass $M_{\rm B}=8.6\,M_{\odot}$, this would imply a companion mass $2\lesssim M_{2}/M_{\odot} \lesssim 3$ for inclinations ranging from 45 to 90 degrees (encompassing 71\% of randomly oriented orbits). However, \citet{Lennon2021} infer a nearly face-on inclination, $i\lesssim 9$ deg, and on this basis estimate a companion mass $M_2 \gtrsim 25\,M_{\odot}$. Since no 25\,$M_{\odot}$ luminous companion is detected spectroscopically, they conclude that the companion is a BH. 

This nearly face-on inclination is inferred under the assumption that the narrow-lined B star's rotation is tidally synchronized and aligned with the orbital plane, such that $2\pi R_{{\rm B}}/P_{{\rm orb}}=\left(v\sin i\right)/\sin i$. Because the projected rotation velocity $v\,\sin i$ is small (\citealt{Lennon2021} estimate $v\,\sin i = 10\,\rm km\,s^{-1}$, with a firm upper limit of $15\,\rm km\,s^{-1}$), the orbit would have to be viewed nearly face-on to reconcile the measured projected rotation velocity with the much larger expected total rotation velocity, $2\pi R_{\rm B}/P_{\rm orb}\approx 100\,\rm km\,s^{-1}$.

No X-rays are detected from the system, and an upper limit of $L_X < 1.4 L_{\odot}$ is inferred in the XMM-Newton 0.2-12 keV band. This is smaller by a factor of 10-100 than the expected accretion luminosity if the BH were accreting a significant fraction of the wind of a close $8.6\,M_{\odot}$ main-sequence companion, but it does not strictly rule out a BH companion because accretion may be radiatively inefficient at the low relevant accretion rates, $\dot{M}\lesssim 10^{-11}\,M_{\odot}\,\rm yr^{-1}$ \citep[e.g.][]{Narayan1995}.

A second B-type star -- one which rotates rapidly, with projected rotation velocity $v\sin i\approx 300\,\rm km\,s^{-1}$ -- is also clearly detected in the spectra of NGC 2004 \#115. Strong Balmer emission lines are found in some of the observed spectra, suggesting that the rapidly rotating star is a Be star.  However, \citet{Lennon2021} found that this star does not seem to display RV-shifts in anti-phase with the narrow-lined B star. In addition, they found that the narrow-lined star's orbit was not well-fit by a simple Keplerian orbit over long timescales. On these grounds, they concluded that the system is a triple, with the narrow-lined star orbiting a dark companion, and the rapidly-rotating star an outer tertiary. The narrow- and broad-lined stars appear to contribute roughly equal fractions of the optical flux. 

\citet{Lennon2021} analysed a light curve of the object from the MACHO survey \citep{Alcock2000} and did not detect evidence of ellipsoidal variability or other variability associated with the orbital period. They interpreted this, together with a low reported flux uncertainty in {\it Gaia} data for the source, as further evidence that the inner orbit is viewed nearly face-on, since the expected ellipsoidal variability amplitude decreases at low inclination. 

If NGC 2004 \#115 indeed contains a 25\,$M_{\odot}$ BH, it would be an extraordinary system. The BH would would be among the most massive electromagnetically detected stellar-mass BHs, and the close orbit with a massive star would likely make the system an ultraluminous X-ray source within 10-20 Myr \citep[e.g.][]{Mondal2020}. Moreover, the entire triple system today would have likely fit within the radius of the BH's progenitor when it was a luminous supergiant, meaning that a remarkable evolutionary history would be needed to keep the system dynamically stable through the BH's formation and to the present day. It is thus worth investigating the system in more detail.

We present a different model for the system: one that contains three luminous stars viewed at an intermediate inclination, with a B star whose rotation is not synchronized. The remainder of this paper is organized as follows. Section~\ref{sec:RVs} discusses our analysis of the spectra and radial velocities of the system. Section~\ref{sec:phot} is focused on photometry and our modeling of the observed light curve variability. We discuss possible interpretations of the system in Section~\ref{sec:discussion}, and conclude with a few general notes of caution regarding the interpretation of non-accreting BH candidates. Appendix~\ref{sec:rvdetails} provides additional details about our measurement of radial velocities (RVs).

\section{Spectral analysis}
\label{sec:RVs}
We retrieved the sky-subtracted, barycenter corrected VLT/FLAMES spectra of NGC 2004 \#115 from the project website.\footnote{https://star.pst.qub.ac.uk/~sjs/flames/datarelease/lmc/index.html} The data are described in detail by \citet{Evans2006}. In total, there are 34 usable spectra, with a typical spectral resolution $R\approx 20,000$ and per-pixel signal-to-noise ratio of $\approx 30$. Six different grating settings were used, with (nearly) non-overlapping wavelength coverage.  Each grating was used for two observing blocks, each split into 3 consecutive exposures separated by $\approx$40 minutes. As a result, the RV of the B star typically varies by only a few $\rm km\,s^{-1}$ within each observing block, but by up to $\approx$140\,$\rm km\,s^{-1}$ across observing blocks.

\subsection{Atmospheric parameters}

Two luminous components are clearly detected in the spectra: a narrow-lined B star that is obviously RV-variable, and a rapidly-rotating Be star whose RV variations are more ambiguous. We modeled the spectra as a sum of two TLUSTY/SYNSPEC models \citep{Hubeny_1995, Hubeny_2011} from the LMC-metallicity BSTAR06 grid \citep[][]{Lanz_2007}. We find atmospheric parameters, abundances, rotation velocities, and luminosity ratios broadly consistent with the values reported by \citet{Lennon2021}, and so for consistency, we adopt their reported values in the rest of our analysis. In brief, the narrow-lined B star contributes about 58\% of the $V$-band light, has temperature $T_{{\rm eff}, B} \approx 22.6\,\rm kK$, surface gravity $\log\left[g/\left({\rm cm\,s^{-2}}\right)\right]\approx 4.0$, an unusually low projected rotation velocity, $v\sin i \lesssim 10\,\rm km\,s^{-1}$, and typical abundances for the LMC. The $v\,\sin i$ value is close to the instrumental resolution of the spectra, so the true value could be even lower. The Be star has $T_{\rm eff,Be}\approx 20\,\rm kK$, $\log\left[g/\left({\rm cm\,s^{-2}}\right)\right]\approx 4.2$, and $v\sin i \approx 300\,\rm km\,s^{-1}$. Double-peaked emission lines are obvious in H$\alpha$ but contribute only subtle infilling to the high-order Balmer lines.

\subsection{Radial velocities}

Although we find consistent atmospheric parameters to \citet{Lennon2021}, we do not find consistent RVs for all epochs. This is shown in detail in Appendix~\ref{sec:rv_comparison}. We measure RVs for the B star and Be star simultaneously by cross-correlating the appropriately normalized TLUSTY/SYNSPEC model spectra with the data, considering several different regions of each grating separately to estimate realistic RV uncertainties, and masking regions containing sky lines, cosmic rays strikes, and interstellar absorption. While we find RVs for the B star that are consistent with those reported by \citet{Lennon2021} within $5\,\rm km\,s^{-1}$ for most epochs, there are a few epochs in which the disagreement is of order 20\,$\rm km\,s^{-1}$ ($20-40$ sigma) and the RV reported by \citet{Lennon2021} appears to be clearly inconsistent with the data (e.g. Figure~\ref{fig:rvfit_compariosn}). The largest discrepancies are in the spectra obtained with the ``HR 02'' grating centered on 3958\,\,\AA. 

We tried fitting the B star's measured RVs with both binary and hierarchical triple solutions. As described in \citet{El-Badry2018}, we first search for solutions using simulated annealing and subsequently initiate a Markov chain Monte Carlo exploration of the posterior in the vicinity of the preliminary solutions \citep{emcee2013}. When fitting a hierarchical triple solution, we model the orbit as a sum of two binary solutions. That is, we fit the standard Keplerian orbital parameters for an inner and outer orbit simultaneously and sum the predicted RVs. 

\begin{figure}
    \centering
    \includegraphics[width=\columnwidth]{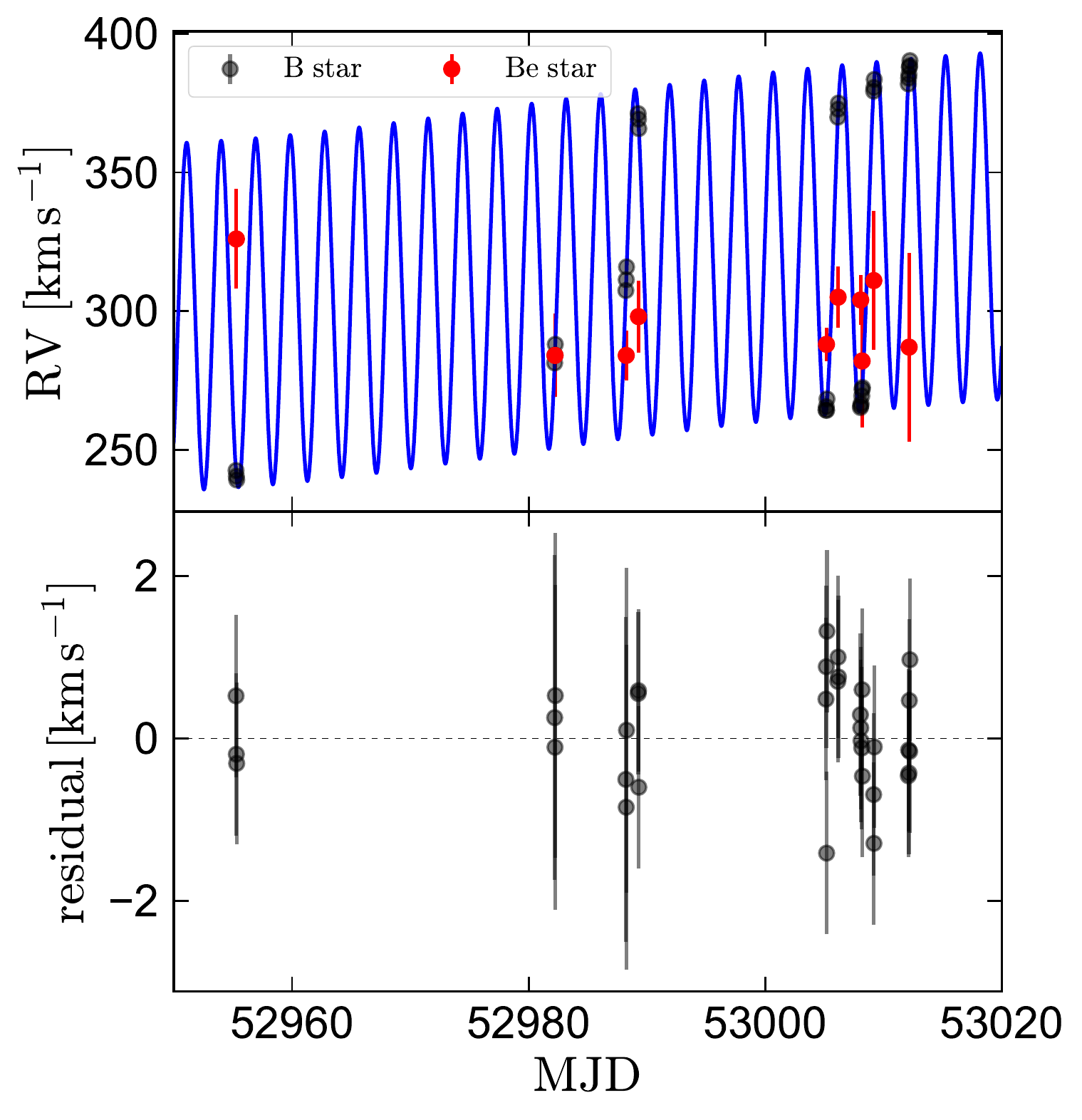}
    \caption{Radial velocities. Black points show the narrow-lined B star; red points show rapidly-rotating Be star, which we assume is an outer tertiary. Blue line shows a hierarchical triple fit to the RVs of the B star. }
    \label{fig:rvs}
\end{figure}

\begin{figure*}
    \centering
    \includegraphics[width=\textwidth]{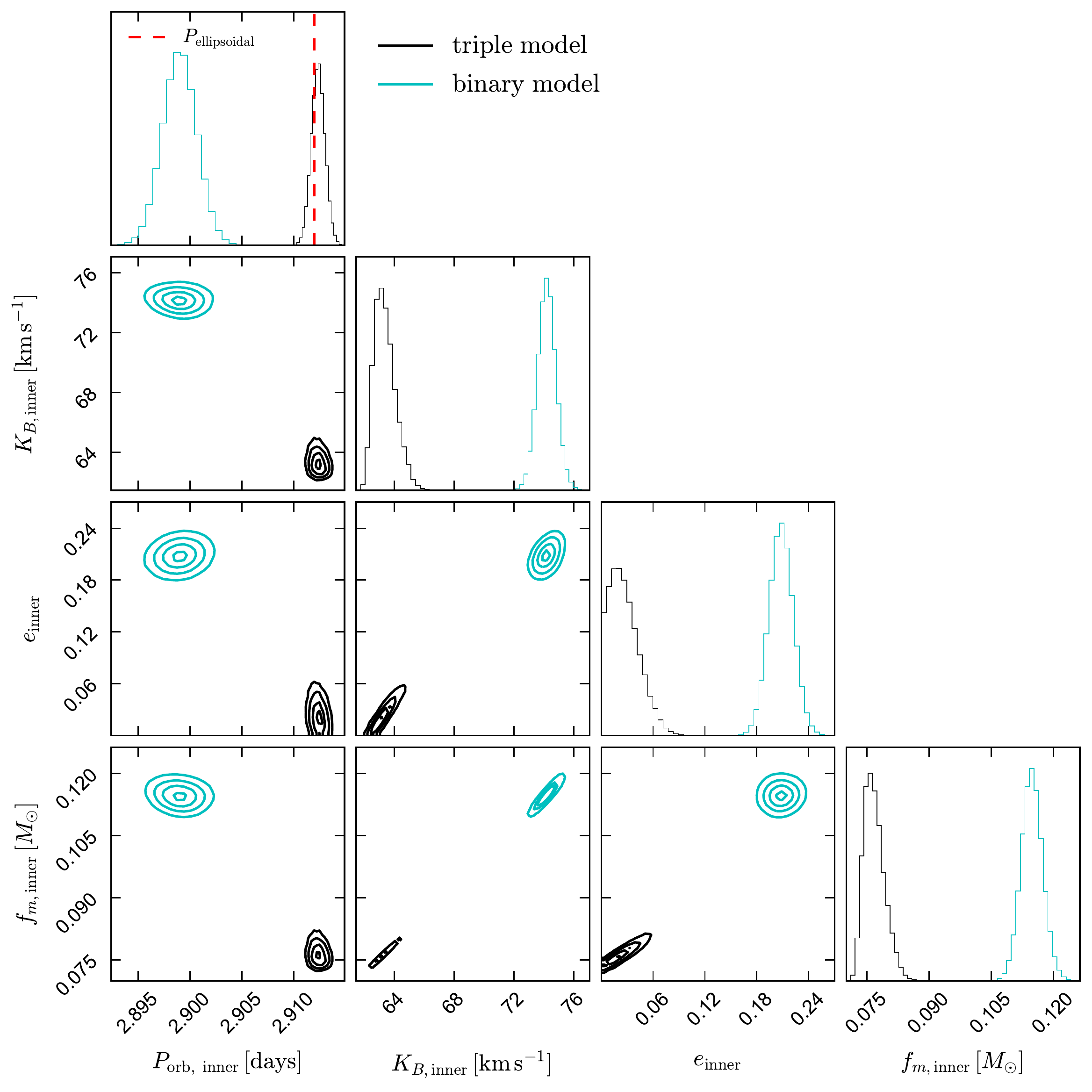}
    \caption{Comparison of RV solution parameters for the B star when its orbit is modeled as a binary (cyan) and as the inner subsystem in a hierarchical triple (black). A binary orbit requires a non-zero eccentricity and an orbital period that is strongly inconsistent with the photometric period (red dashed line). In the triple model, the eccentricity of the inner orbit is consistent with 0 and the orbital period matches the photometric period. The mass function is somewhat lower in the triple solution, but is relatively low in both cases.}
    \label{fig:corn}
\end{figure*}

The results of this fitting are explored in Figures~\ref{fig:rvs},~\ref{fig:corn}, and~\ref{fig:rvs_comparisons}. We obtain a good fit with a triple solution, which is shown in Figure~\ref{fig:rvs}. This solution has a circular (or at least nearly circular; $e<0.04$) inner orbit, and a period that matches the photometric period (Section~\ref{sec:phot}).  We also found a marginally acceptable solution when fitting a pure binary orbit (Figure~\ref{fig:rvs_comparisons}), which has significant eccentricity, $e\approx 0.21$. The fit is formally worse than the triple solution ($\Delta \chi^2 = 24$) but should not be immediately discounted on this basis, since the triple solution has several additional free parameters. However, the orbital period in the binary solution is inconsistent with the photometric period (Figure~\ref{fig:corn}), and forcing the period to match the photometric value yields an obviously unsatisfactory fit (Figure~\ref{fig:rvs_comparisons}). This, together with evolutionary considerations (Section~\ref{sec:discussion}) leads us to reject the binary solution and model the system as a triple. This is the same broad conclusion reached by \citet{Lennon2021}, although some of their RV measurements were significantly different and they did not detect a photometric period. Despite the discrepant RV measurements, \citet{Lennon2021} find a qualitatively similar solution for the inner binary to ours. They do this fitting a pure binary solution to only a subset of the RVs obtained within a short ($\sim1$ week) period, within which the effects of the tertiary are minimal. The RVs they measure for these epochs agree with ours within a few $\rm km\,s^{-1}$.

While the parameters of the inner binary are well-constrained (Table~\ref{tab:summary}), those of the outer orbit are not. We can confidently rule out an outer period shorter than 100 days, but the RV semi-amplitude, eccentricity, and orientation of the outer orbit, as well as the center-of-mass velocity of the system, are still uncertain. That is, any outer period above 100 days can reproduce the linear trend in the inner orbit's center-of-mass when combined with suitable outer semi-amplitude and eccentricity. Given our constraints on component masses (Section~\ref{sec:discussion}) and the fact that we observe a $\sim25\,{\rm km\,s^{-1}}$ change in the inner binary's center-of-mass RV over a $\sim60$ day baseline, a conservative upper limit on the outer period is $P_{{\rm outer}}<1000\,{\rm days}$ (or $P_{{\rm outer}}<400\,{\rm days}$ for a circular orbit) , with a value of a few hundred days most likely. This corresponds to an outer semi-major axis of a few au. 

The Be star appears to move in anti-phase with the center-of-mass of the inner binary (Figure~\ref{fig:rvs}), suggesting that it is indeed the outer tertiary. However, the observational uncertainties on its RVs are much larger than those on the B star's RVs. We thus do not attempt to model its orbit or dynamically constrain its mass. Longer-term monitoring of the B star's RVs is needed to fully solve the orbit. We do note, however, that the apparent reflex acceleration of the Be star is of comparable magnitude to the RV shift of the inner binary's center of mass. This suggests that the total mass of the inner binary is comparable to the Be star's mass within about a factor of 2. If the inner binary contained a 25\,$M_{\odot}$ BH, its total mass would be a factor of 5 larger than the $6-7\,M_{\odot}$ mass of the Be star inferred by \citet{Lennon2021}. One would then expect an RV shift of the Be star of 150-200\,$\rm km\,s^{-1}$ over the time baseline of the FLAMES data, which is not observed.

\begin{table}
\caption{Orbital constraints from joint fitting of RVs and light curve. Uncertainties are one sigma. Parameters of the inner orbit are well-constrained; those of the outer orbit are not. In fitting the RVs, we ultimately fix the inner period to the photometric value, which is precisely constrained from 27 years of data. $T_0$ is defined such that the B star is farthest from the observer at phase 0; i.e., its maximum blueshift occurs at phase 0.25.}
\label{tab:summary}
\begin{tabular}{llll}
Parameter & Description   & Units & Constraint \\
\hline
$P_{\rm orb,\, inner}$  &  Inner orbital period   &  days   & $ 2.91203\pm 0.00001$ \\
$T_{0,\,\rm inner}$   & Inner orbit time of phase 0  &  days &  $2455267.42 \pm 0.02$        \\
$e_{\rm inner}$    & Inner orbit eccentricity &   --   &  $<0.04$   \\
$K_{B,\rm inner}$    & B star inner semi-amplitude & $\rm km\,s^{-1}$ & $62.6\pm 0.5$           \\
$f_{m,\,{\rm inner}}$       & Inner mass function &  $M_{\odot}$   &  $0.074\pm 0.002$         \\
\hline
\end{tabular}
\end{table}
 
\section{Photometric variability}
\label{sec:phot}

To search for evidence of photometric variability associated with binarity, we analyzed light curves obtained by the MACHO and OGLE projects. 

\begin{figure*}
    \centering
    \includegraphics[width=\textwidth]{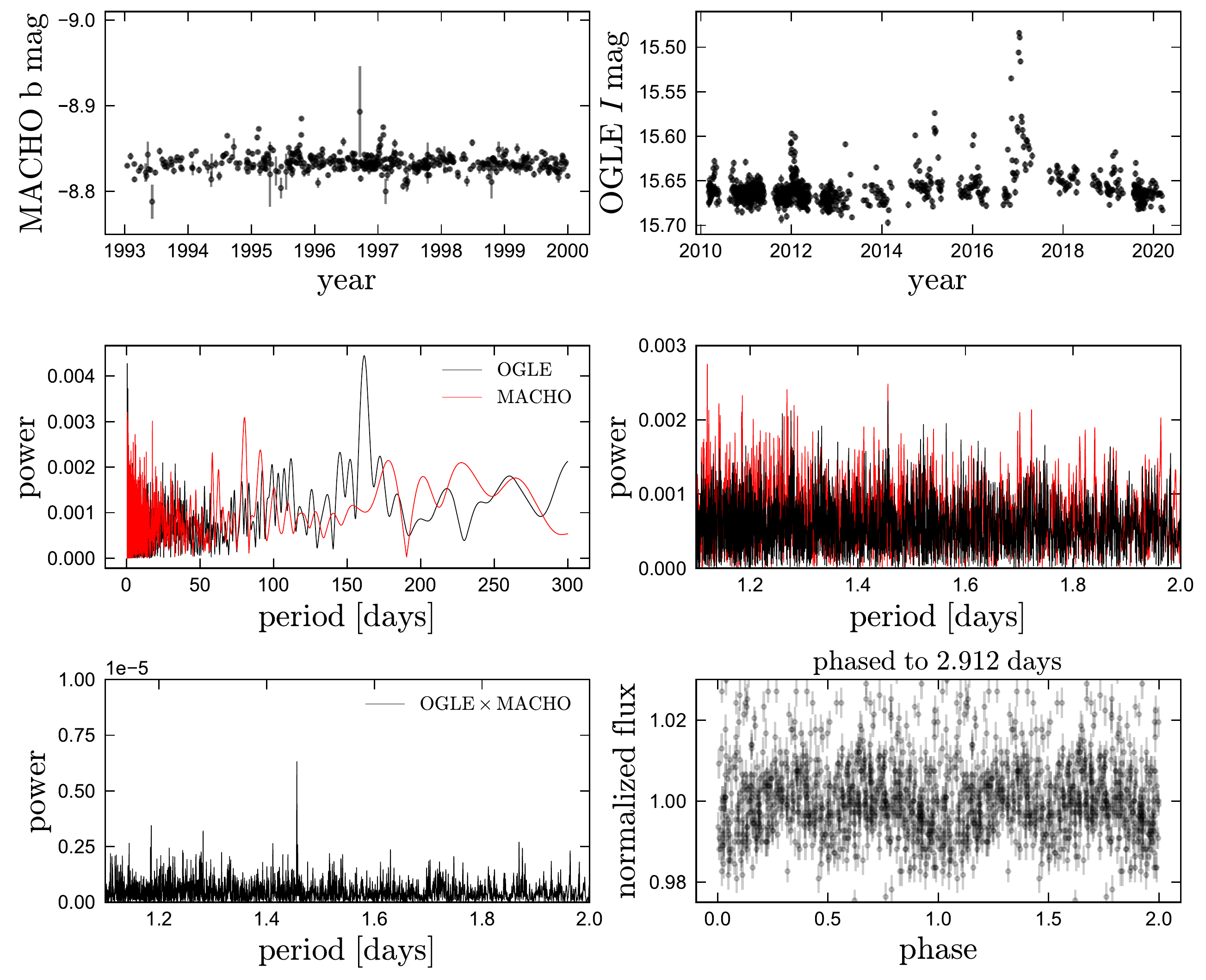}
    \caption{Photometric analysis. Top panels show light curves of NGC 2004 \#115 from MACHO (left) and OGLE (right). On long timescales, the source shows irregular variability with frequent brightening episodes of a few percent, likely due to changes in the Be star's disk. Middle panels show Lomb-Scargle periodograms of both datasets, with the right panel zoomed in on short periods. Several marginally significantly peaks are apparent, but only the peak at $P=1.456$ days is present in both light curves. This illustrated in the bottom left panel, which shows the product of the two periodograms. Bottom right panel shows the OGLE light curve, phased to twice this period, which is the orbital period of the inner binary. }
    \label{fig:phot_summary}
\end{figure*}

\begin{figure*}
    \centering
    \includegraphics[width=\textwidth]{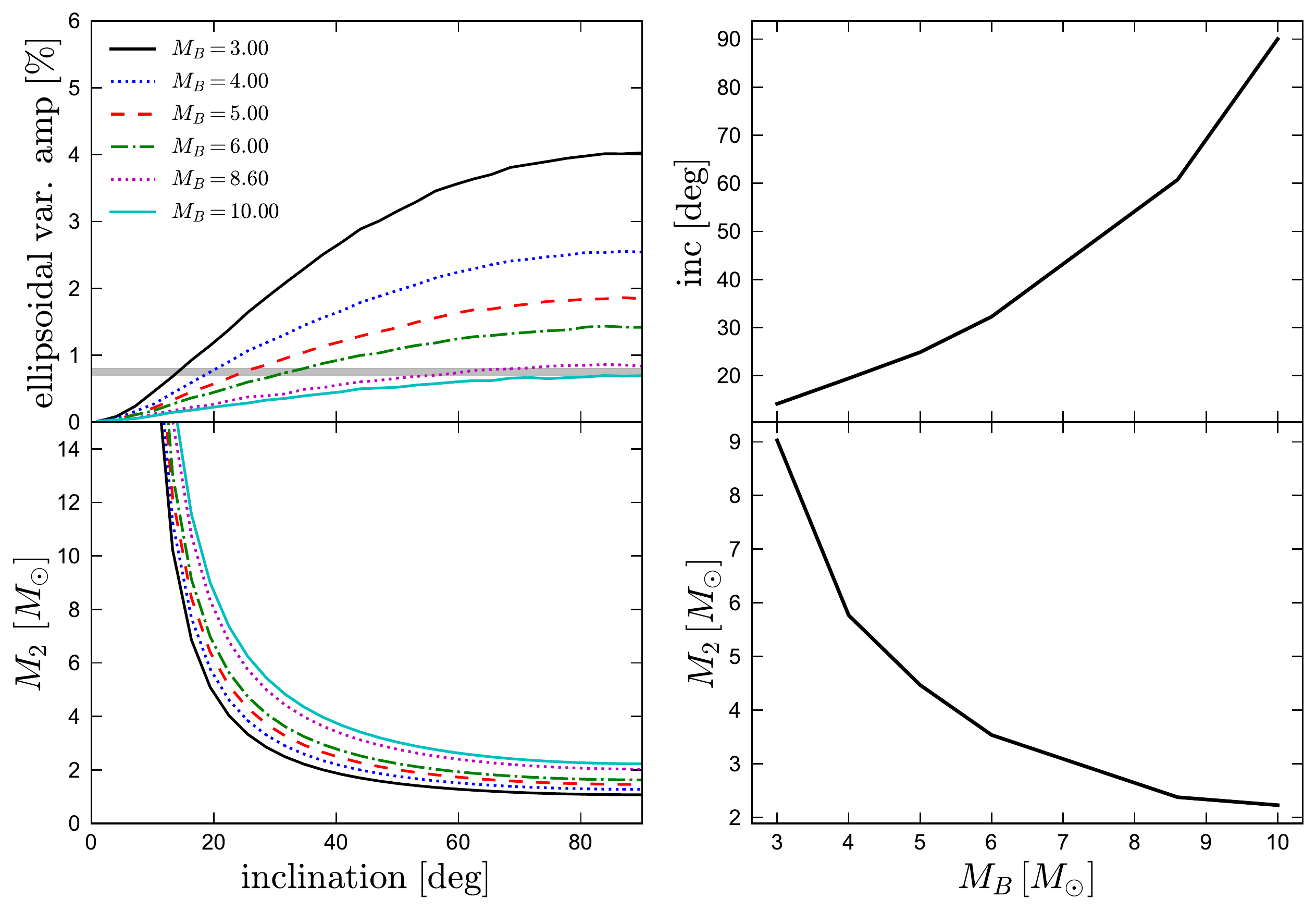}
    \caption{Predicted ellipsoidal variability amplitude (peak-to-peak, including dilution from the tertiary) and implied companion mass as a function of inclination and mass of the narrow-lined B star. Gray shaded region in the upper left shows the observed ellipsoidal variability amplitude. Right panels show the inclination and companion mass required to match the observed ellipsoidal variability amplitude. Assuming the narrow-lined B star is a normal main-sequence star with mass of $8-9\,M_{\odot}$, the companion mass is about $2.5\,M_{\odot}$ and the inclination is 50-70 degrees. }
    \label{fig:implications}
\end{figure*}

\subsection{MACHO light curve}
\label{sec:macho}

We analyze the $V-$band light curve of NGC 2004 \#115 obtained by the MACHO project \citep{Alcock2000}. The same data was also analyzed by  \citet{Lennon2021}. The MACHO (field, tile, sequence) identifier is (61, 8312, 25). The light curve has 413 photometric points obtained between 1993 and 1999, with a median reported photometric uncertainty of 0.003 mag. It is  shown in the upper left panel of Figure~\ref{fig:phot_summary}. Magnitudes are instrumental. The MACHO project also obtained an $R-$band light curve for the source, but we do not analyze it because it contains significantly fewer measurements. 


\subsection{OGLE light curve}
\label{sec:OGLE}
We also analyze the calibrated $I$-band light curve of NGC 2004 \#115 obtained by the optical gravitational microlensing experiment \citep[OGLE;][]{Udalski2003, Udalski2015}. The light curve contains 850 photometric measurements taken between 2010 and 2020, with a median reported uncertainty of 0.004 mag, and is shown in the upper right panel of  Figure~\ref{fig:phot_summary}. 

\subsection{Light curve analysis}
Both light curves show evidence of variability on a range of timescales. Particularly in the OGLE data, there are several episodes in which the source brightened rapidly and then faded over a timescale of order a month. Such behavior is very common in Be stars and is usually attributed to growth and fading of their circumstellar disks \citep[e.g.][]{Rimulo2018}. The amplitude of long-term variability is low compared to typical Be stars. This owes at least in part to dilution by the B star, which contributes more than half the total light. There is also variability on short timescales: the scatter between adjacent light curve points is larger than expected due to photometric uncertainties alone. 

The middle panels of Figure~\ref{fig:phot_summary} show Lomb-Scargle periodograms of both light curves. A number of peaks are evident in both periodograms, but it is not immediately obvious which (if any) are significant. To identify periods that are present in both light curves and thus less likely to be due to correlated noise, we multiplied the two periodograms (bottom left panel). The strongest surviving peak (besides the one at 1 sidereal day) is at 1.456 days. This is almost exactly half the spectroscopically measured orbital period of the inner binary, as is expected for ellipsoidal variability due to tidal distortion. In the bottom right panel of Figure~\ref{fig:phot_summary}, we thus show the OGLE light curve phased to twice this period. Clear quasi-sinusoidal variability is apparent, with a deeper minimum at phase 0 than at phase 0.5. As we show in Section~\ref{sec:ellip_reflect}, this is a consequence of the combined effects of ellipsoidal variation and reflection.

The observed ellipsoidal variability amplitude provides a joint constraint on the companion mass and inclination, essentially solving the system if the mass of the B star is known. This is illustrated in Figure~\ref{fig:implications}. For a given B star mass and inclination, we solve for the inner companion mass $M_2$ required to reproduce the observed mass function. We then calculate the predicted peak-to-peak ellipsoidal variability amplitude using PHOEBE \citep[][]{Prsa2005}, a code for modeling binary star light curves. Following \citet{Lennon2021}, we assume $R_{B}=5.6\,R_{\odot}$ and $T_{\rm eff,B}=22.6\,\rm kK$, and we assume the B star contributes 58\% of the total light in the system. Here we model the companion as dark; i.e., we do not consider reflection or eclipses. 

The peak-to-peak amplitude of ellipsoidal variability alone in the observed light curve is 0.75\%. This is shown with the gray shaded region in Figure~\ref{fig:implications}. For a given $M_{B}$, only a small range of $M_2$ and inclination lead to a predicted ellipsoidal variability amplitude consistent with this observed value. This is shown explicitly in the right panels of Figure~\ref{fig:implications}, where we plot the combination of inclination and companion mass that would reproduce both the observed ellipsoidal variability amplitude and the inner binary's mass function for a given $M_{B}$. Perhaps counter-intuitively, the implied mass of the companion is anticorrelated with $M_B$. This is because a lower $M_B$ implies a larger Roche lobe filling factor for the B star, and thus requires a lower inclination to match the observed ellipsoidal variability amplitude. 

The observed variability cleanly rules out a $25\,M_{\odot}$ BH companion, which would produce an ellipsoidal variability amplitude of only 0.08\%. The reason for this is that the amplitude of ellipsoidal variation scales roughly as $\sin^2 i$ at fixed Roche lobe filling factor, so in the BH scenario it is reduced from its edge-on  value by a factor of $\approx \sin^2 (9^{\circ})\approx 0.024$. For a lower-mass companion, the implied inclination is higher. We emphasize that these calculations do not depend on any assumptions about whether the B star is rotating synchronously: at the low rotation rates relevant for this system, rotation has negligible effects on the light curve.

\subsection{Light curve model}
\label{sec:ellip_reflect}
\begin{figure}
    \centering
    \includegraphics[width=\columnwidth]{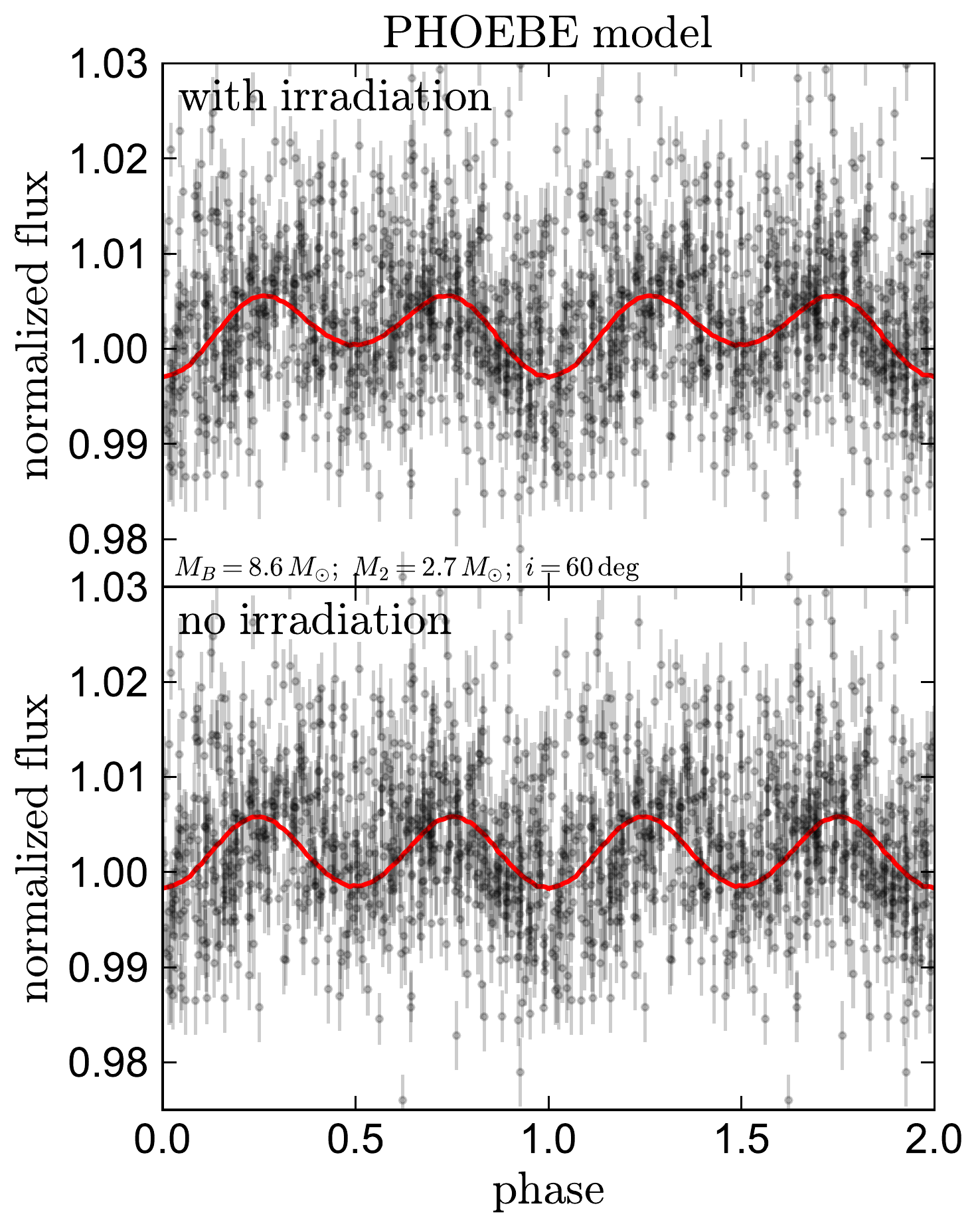}
    \caption{PHOEBE light curve model for the inner binary. Top panel shows a model with a luminous, 2.7\,$M_{\odot}$  main-sequence companion. The light curve minimum at phase 1 is significantly deeper than the one at phase 0.5 due to the reflection/irradiation of the secondary. I.e., the source is brighter when the side of the secondary that is heated by the B star faces the observer. Bottom panel shows the same model but with a dark, point-like secondary. In this case, adjacent minima have (nearly) equal depth, yielding a worse match to the data. This detection of a reflection effect provides unambiguous evidence of a cooler, luminous-star secondary. }
    \label{fig:phoebe}
\end{figure}

We used PHOEBE to predict light curves of the inner binary for a range of companion masses and inclinations. The primary result of this analysis is shown in Figure~\ref{fig:phoebe}, which compares the OGLE data to a PHOEBE model for an $8.6\,M_{\odot}$ primary and a $2.7\,M_{\odot}$ main-sequence secondary (including dilution from the Be star), with $i=60$ deg. For the secondary, we assume $T_{\rm eff,2}=12500\,\rm K$ and $R_2 = 1.7\,R_{\odot}$, appropriate for a LMC-metallicity star near the zero-age main sequence. Both stars are assumed to have radiative envelopes and correspondingly, a bolometric gravity darkening exponent $\beta =1$. We prewhiten the plotted OGLE light curve by removing variability with frequency associated with 1 sidereal day.  

The PHOEBE light curve reproduces the morphology of the OGLE light curve when we assume a luminous secondary (top panel of Figure~\ref{fig:phoebe}). Most importantly, there is a significant difference between the depths of adjacent minima, as also found in the data. This asymmetry is a result of the fact that the side of the cool secondary that faces the B star is hotter than the side facing away from the B star, causing the unresolved source to be brighter when this side faces the observer (phase 0.5) than when it faces away (phase 1). The asymmetry is absent when we model a dark secondary (bottom panel of Figure~\ref{fig:phoebe}). This detection confirms that the companion is luminous rather unambiguously, as a detached BH or neutron star would lack a sufficiently large surface to produce a detectable reflection effect.  \footnote{We note that unequal minima can also be produced by gravity darkening in near-Roche lobe filling ellipsoidal variables. We tested whether the observed asymmetry can be reproduced by a Roche lobe-filling B star with a dark companion viewed at low inclination, but we find that it is weaker in this case than observed.} We also checked that the phasing inferred from the light curve modeling is consistent with that inferred from fitting the RVs. The two values of $T_0$ are consistent to within 0.03 days.\footnote{We  note that changes in light travel time due to the outer orbit are expected to ``blur'' the phased photometry somewhat. We do not attempt to model this because the expected shifts are short ($\lesssim 1$ hour for plausible outer orbits, or $\lesssim 1$\% of the inner orbit), the light curve contains no sharp features, and the signal-to-noise ratio is too low to reliably phase subsets of the light curve to the required precision.} The formal fit of both models is relatively poor (reduced $\chi^2 \approx 8$) because the light curve contains significant irregular variability driven by outbursts from the Be star. The model with irradiation does, however, provide a significantly better fit than the model with a dark companion ($\Delta \chi^2 = 130$). 

The amplitude of the reflection effect allows us to place an additional constraint on the mass of the B star, because the amplitude of the reflection effect depends on the temperature difference between the two stars. For example, while the ellipsoidal amplitude and RVs could be reproduced with $M_{\rm B}\approx 4 M_{\odot}$, $M_2 \approx 6 M_{\odot}$. and $i\sim 20$ deg, the reflection effect amplitude in this case would be too low to be consistent with the data, because a 6 $M_{\odot}$ main-sequence companion would have a temperature too similar to the B star to cause as large a reflection effect as is observed. This is further evidence against a scenario in which the Be star is the secondary.

A significantly larger B star mass can also be ruled out. For $M_B\gtrsim 9\,M_{\odot}$, the implied inclination is large enough ($i \gtrsim 69$ deg) that eclipses would be detected in the light curve. Our joint analysis of the light curve and RVs thus limits the B star's mass to $5\lesssim M_B/M_{\odot} \lesssim 9$, and the companion mass to $2.3 \lesssim M_2/M_{\odot} \lesssim 4.5$, irrespective of the evolutionary state of the B star. 

\subsection{Is the observed ellipsoidal variability inconsistent with the reported {\it Gaia} flux uncertainty?  }
\citet{Lennon2021} attempt to constrain NGC 2004 \#115's photometric variability based on its reported {\it Gaia} flux error, and thus reported a peak-to-peak ellipsoidal variability upper limit of 0.5\%, which is inconsistent with the OGLE data. We believe this is a consequence of misinterpreting the reported {\it Gaia} flux uncertainties.

In particular, \citet{Lennon2021} interpret the reported \texttt{phot\_g\_mean\_flux\_error/phot\_g\_mean\_flux} = 0.001 as meaning that the source's  brightness is constant across all scans at the 0.1\% level. However, the parameter \texttt{phot\_g\_mean\_flux\_error} represents the standard deviation of the single-epoch $G$-band fluxes {\it divided by the square root of the number of visits}, which is reported as \texttt{phot\_g\_n\_obs}. Thus a more reliable estimate of a source's RMS flux variability across scans is 
\begin{equation}
    \label{eq:varindex}
    V_{G}=\frac{\rm phot\_g\_mean\_flux\_error}{\rm phot\_g\_mean\_flux }\sqrt{\rm phot\_g\_n\_obs},
\end{equation}
where the column names are from the {\it Gaia} archive \citep[e.g.][]{El-Badry2021_cvs}. For NGC 2004 \#115, $V_G\approx 0.019$, corresponding approximately to a 1.9\% RMS flux variability. This is consistent with the variability found in the OGLE and MACHO light curves, and larger than the limit assumed by \citet{Lennon2021} by a factor of $\sqrt{\rm phot\_g\_n\_obs}\approx 21$.

$V_G$ not a measurement of purely astrophysical variability, since some variability is due to photon noise and photometric systematics (see \citealt{Guidry2020} for detailed discussion). However, the median $V_G$ for sources with $G\sim 15.5$ is 0.006, and the measured $V_G=0.019$ makes NGC 2004\,\#115 more variable than 94\% of all {\it Gaia} sources with similar $G$-band magnitude. This implies that the source is genuinely astrophysically variable at the 1-2\% level, as also shown by the MACHO and OGLE light curves. 

\section{Summary and Discussion}
\label{sec:discussion}

\begin{figure*}
    \centering
    \includegraphics[width=\textwidth]{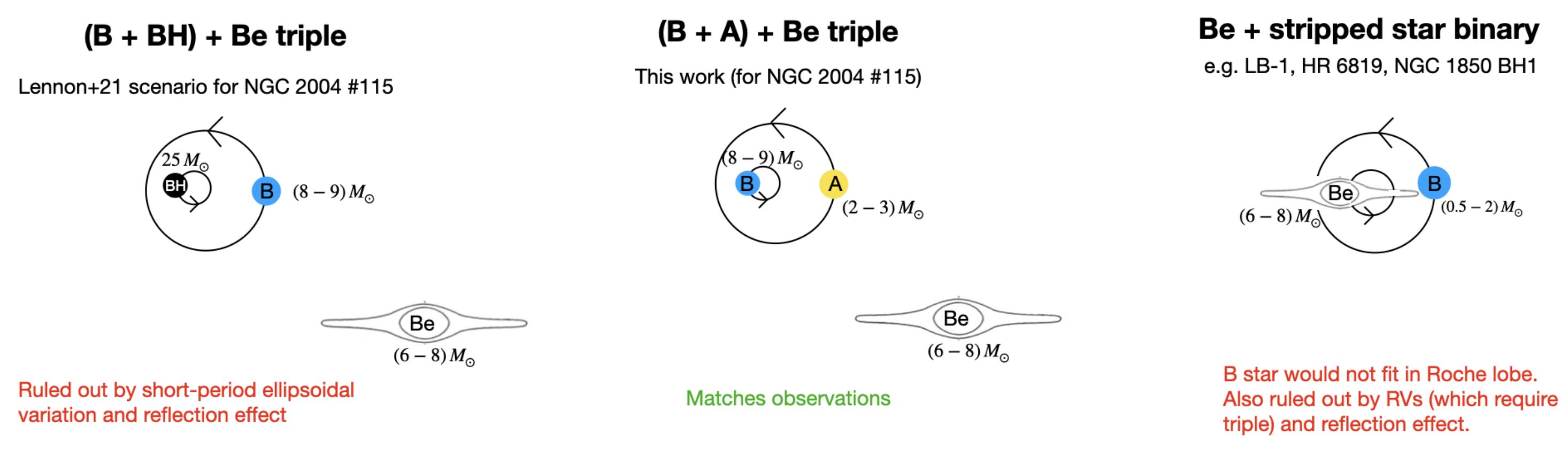}
    \caption{Cartoon summary of the models we consider for NGC 2004 \#115. The model proposed by \citet{Lennon2021}, which contains an inner BH + B star binary and outer Be star (left), is ruled out by the observed ellipsoidal variations and reflection effect (Figures~\ref{fig:implications} and~\ref{fig:phoebe}). Our preferred  model (center) replaces the BH with a $2-3\,M_{\odot}$ A-type star. A stripped star + Be star binary model, as has been successfully invoked for several similar systems, is ruled out for NGC 2004 \#115 by both the amplitude of ellipsoidal distortion (Figure~\ref{fig:roche}) and the evidence for the tertiary in the B star's RVs (Figure~\ref{fig:rvs}).  }
    \label{fig:schematic}
\end{figure*}

We have detected photometric variability in NGC 2004\,\#115 that is associated with the orbital period of the inner binary and which we interpret as a combination of ellipsoidal variation (i.e., tidal deformation of the B star) and reflection (i.e., heating of one side of a cooler, luminous companion; see Figure~\ref{fig:phoebe}). Detecting the unseen companion spectroscopically would be quite challenging, as it is expected to contribute less than 1\% of the total luminosity.

We considered three possible scenarios for the system, as illustrated schematically in Figure~\ref{fig:schematic}. We find that three luminous stars are required to reproduce the data. 

\subsection{Is a stripped star scenario possible?}
NGC 2004 \#115 is similar in several aspects to the BH candidates  LB-1 \citep{Liu2019}, HR 6189 \citep{Rivinius2020}, and NGC 1850 BH1 \citep{Saracino2021}. Given that these systems can be well described by models in which an undermassive stripped star orbits a Be star \citep[e.g.][]{Shenar2020, Bodensteiner2020, El-Badry2021, El-Badry2021_burdge}, it is worth considering carefully whether such a scenario could also explain NGC 2004 \#115. An attractive feature of the stripped star scenario is that it simultaneously explains the slow rotation of the B star and the rapid rotation of the Be star. 

\begin{figure}
    \centering
    \includegraphics[width=\columnwidth]{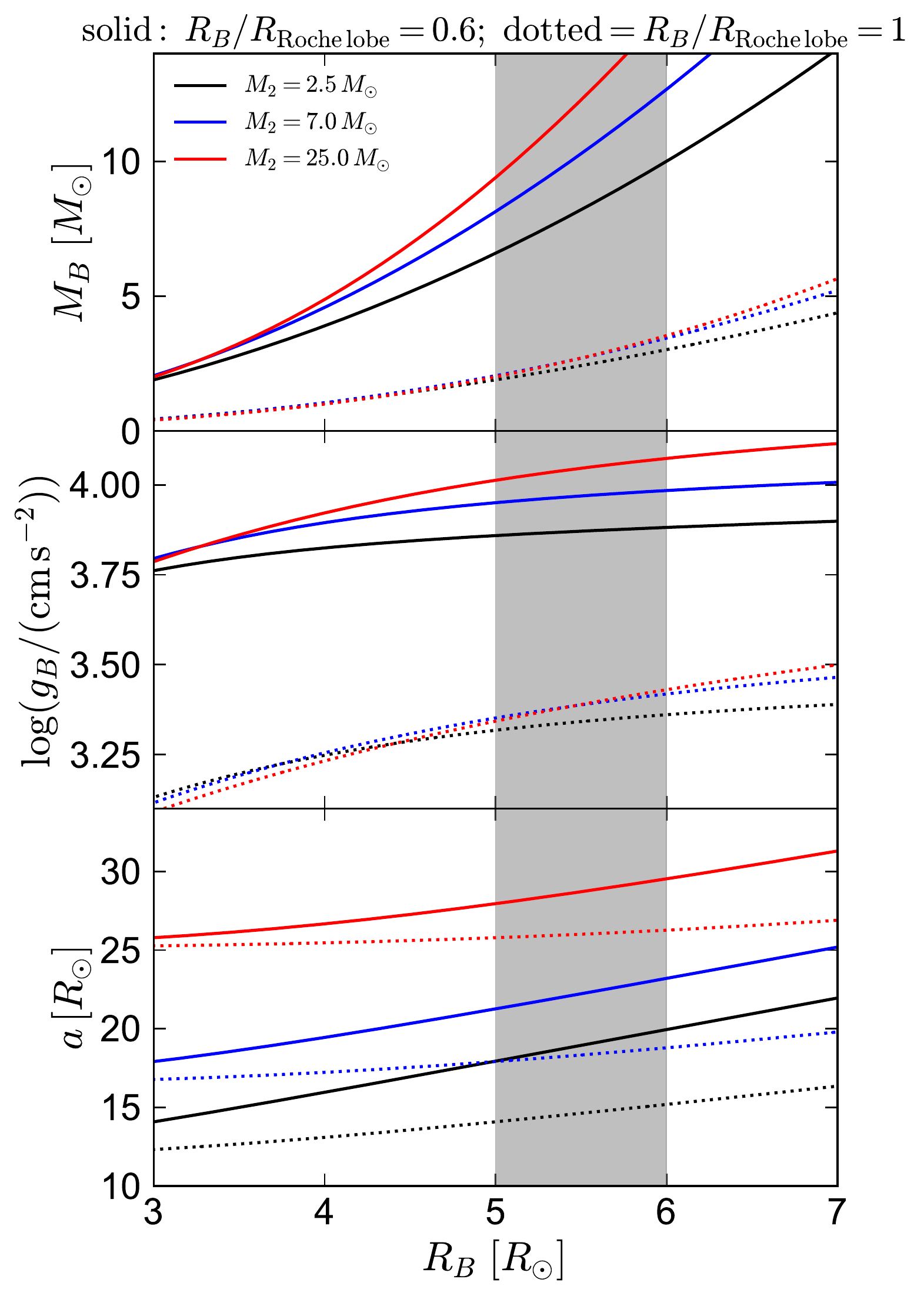}
    \caption{B star mass and surface gravity, and binary semi-major axis predicted for different B star radii if the Roche lobe filling factor is 60\% (solid lines) or 100\% (dotted lines). Shaded region shows the constraint on $R_{\rm B}$ from the spectroscopic temperature, distance, and SED. For a 60\% Roche lobe filling factor, a companion mass $M_2 \approx 2.5 M_{\odot}$ implies that the B star's mass is between 7 and 10 $M_{\odot}$; i.e., a normal B star on or near the main sequence. The lowest possible mass of the B star, if it fills its Roche lobe and the system were viewed nearly face on, would be $2\,M_{\odot}$. In this case, the surface gravity would be significantly lower, $\log g\approx 3.3$. An even lower-mass B star would overflow its Roche lobe for any companion mass.}
    \label{fig:roche}
\end{figure}

There are, however, several serious problems with such a scenario in NGC 2004\,\#115: 

\begin{itemize}
    \item The RVs of the B star are poorly fit by a pure binary model, particularly when we require the orbital period to match the photometric period (Figure~\ref{fig:rvs_comparisons}). Modeling the system as a triple solves this issue (Figure~\ref{fig:corn}). 
    \item The surface abundances of the B star are normal for a star in the LMC \citep{Lennon2021}. In the stripped star scenario, significant enhancement of helium and nitrogen, and depletion of carbon and oxygen, are expected.
    \item The B star's mass must be larger than $\sim 2\,M_{\odot}$ in order for it to fit within its Roche lobe given its observed radius (Figure~\ref{fig:roche}). A stripped star of this mass could be produced from a star that was initially $\sim 10\,M_{\odot}$. \citep[e.g.][]{Gotberg2017} but its current state would be short-lived ($\lesssim 50,000$ years), and its predicted effective temperature at the current radius would be hotter than observed.

    \item The mass of the Be star is likely about 7\,$M_{\odot}$. If the B star were a stripped star with mass $\sim 2.5 M_{\odot}$, the inclination would have to be very low ($\sim 15$ deg) to match the mass function. But the Be star's high projected rotation velocity, $v\,\sin i \approx 300\,\rm km\,s^{-1}$, would be inconsistent with such a low inclination unless its rotation axis were misaligned with the orbital plane, which seems unlikely in a close binary.
    \item The observed reflection effect in the light curve is not well explained in a binary scenario, because the Be star's temperature is similar to that of the B star. 
\end{itemize}

These considerations lead us to conclude that system is a hierarchical triple system with an outer Be star and a slowly rotating B star in the inner binary. The outer period is not well constrained. The OGLE light curve periodogram contains a formally significant peak at a 162 day period (Figure~\ref{fig:phot_summary}). It is tempting to associate this with the outer period, but since no similar peak is found in the MACHO data, further spectroscopic monitoring is needed to determine the outer period.

The triple scenario does not directly explain either why the Be star is rapidly rotating or why the B star is slowly rotating. At least one other well-studied triple system with an outer Be star is known: the 4th magnitude star $\nu$ Gem \citep{Klement2021}. It is possible that the system began as a quadruple and the Be star was spun up by mass transfer from its own unseen companion, or perhaps the Be star is rapidly rotating for another reason. In any case, about a quarter of early B stars in NGC 2004 are Be stars \citep{Martayan2006}, so a Be star tertiary is not as unlikely as it sounds.

\subsection{Should the B star be tidally synchronized?}
\label{sec:synchronization}

In the \citet{Zahn1977} theory, tidal synchronization in stars with radiative envelopes occurs via radiative damping of the dynamical tide. The predicted synchronization timescale in this theory is 

\begin{equation}
    \label{eq:tsync}
    t_{{\rm sync}}=\frac{\beta}{5\times2^{5/3}E_{2}}\left(\frac{R^{3}}{GM}\right)^{1/2}\frac{1}{q^{2}\left(1+q\right)^{5/6}}\left(\frac{a}{R}\right)^{17/2}.
\end{equation}
Here $M$ and $R$ are the mass and radius of the star to be synchronized (in this case, the B star), $q=M_2/M$, where $M_2$ is the companion mass, $\beta = I/(MR^2)$, where $I$ is the star's moment of inertia, $a$ is the semi-major axis, and $E_2$ is the tidal torque coefficient, which depends on the structure of the star, particularly the size of the convective core. A reasonable approximation for main-sequence stars \citep[e.g.][]{Yoon2010} is $E_2= 0.04\times (R_{\rm conv}/R)^8$, where $R_{\rm conv}$ is the radius of the convective core. 

We calculated a model for a 20 Myr-old $8.6\,M_{\odot}$ star using MESA \citep{Paxton_2011, Paxton_2013, Paxton_2015, Paxton_2018, Paxton_2019}. From this model, we find $\beta = 0.048$ and $R_{\rm conv}/R\approx 0.112$. Assuming $M_2=2.7\,M_{\odot}$ and $a\approx 19\,R_{\odot}$, Equation~\ref{eq:tsync} yields a present-day synchronization timescale of $t_{\rm sync}=160$\,Myr, much longer than the age of NGC 2004. However, the synchronization timescale would have been shorter earlier in the B star's life, because $R$ increases and $R_{\rm conv}$ decreases over the course of the star's main-sequence evolution. When the same MESA model was on the zero-age main-sequence, it had $\beta = 0.07$ and $R_{\rm conv}/R\approx 0.219$. Taking its radius at the time to have been $3.6\,R_{\odot}$, we find a ZAMS synchronization timescale of 24 Myr -- significantly shorter than the value today, but still comparable to the lifetime of NGC 2004 and presumed age of the system. We thus conclude that it is not obvious that the B star should be tidally synchronized.  

Both the present-day and ZAMS synchronization timescales we calculate above are much longer than the value reported by \citet{Lennon2021}, who found $t_{\rm sync}=0.07$\,Myr using the formulae from \citet{Hurley2002}. Those formulae are nearly identical to Equation~\ref{eq:tsync}, except that they use a simpler approximation for $E_2$. We believe there are two main reasons for the discrepancy between the values of $t_{\rm sync}$ predicted above and those reported by \citet{Lennon2021}. First, they assumed $M_2 \approx 25\,M_{\odot}$, leading to a $q$ value an order of magnitude too large. Also accounting for the larger $a$ in their scenario, this leads to a factor of $\approx$25 underestimate of $t_{\rm sync}$. Second, they assumed a significantly larger value of $E_2$, based on a fitting function for $E_2(M)$ for ZAMS stars from \cite{Hurley2002}, while also using the present-day radius of $5.6\,R_{\odot}$. These two inconsistencies lead to a factor of $\approx 1000$ underestimate of $t_{\rm sync}$. 


Although it is not clear that the B star should be fully synchronized, the system's age is comparable to the ZAMS synchronization timescale, so it still seems surprising that the B star's surface rotation period is so much lower (about a factor of 10, or more if the observed $v\,\sin i$ value is viewed as an upper limit) than the inner binary's orbital period. Even a 2.9 day rotation period would be somewhat shorter than average for a young B star, so the observed surface rotation period $P_{\rm rot}\gtrsim 25\,\rm days$ is somewhat unexpected.\footnote{It is possible that the B star's spin axis is not aligned with the inner orbital plane; e.g., due to precession of the inner orbit driven by the tertiary. In this case, its inclination could be lower and the rotation period could be somewhat shorter. But it very unlikely that it is viewed so face-on as to be synchronized, since only $\sim1$\% of random orientations would have such a low inclination.} We also note that (a) tidal synchronization in stars with radiative envelopes remains an unsolved problem (there are observed systems that are synchronized where Zahn's theory suggests they should not be; e.g., \citealt{Preece2018} ), and (b) although the tidal {\it circularization} timescale in Zahn's theory is about 2 orders of magnitude longer than the predicted synchronization timescale, the inner binary's orbit is close to circular ($e < 0.04$).

Although they are not the norm, slowly rotating stars in close binaries are not unheard of. Some systems in the literature that are similar to NGC 2004\,\#115 include the triple HD 201433, which contains a B star with surface $v\,\sin i \approx 8\,\rm km\,s^{-1}$, and even slower rotation in the interior, in a 3.3 day orbit \citep{Kallinger2017}; KIC 8429450, another triple with a 2.7 day inner period and a $\approx$38 day core rotation period \citep{Li2020}, and KIC 9850387, a binary with a 2.7 day period, surface $v\,\sin i \lesssim 10\,\rm km\,s^{-1}$, and core rotation period of $\approx$190 days \citep{Li2020, Sekaran2021}. Additional examples with slightly longer orbital periods can be found in \citet{Fuller2021}.

The origin of such slow rotation periods is open to debate. The most obvious way for a B star to lose angular momentum is via mass transfer, as expected in the stripped-star scenario \citep[e.g.][]{Heber2009}. But it is unlikely that there was a previous episode of mass transfer in the inner binary in NGC 2004 \#115, since mass transfer would not be expected to end while the primary was still the more massive star, as it is today. An unusually strong magnetic field can also lead to slow rotation via magnetic braking, but there is no evidence of an unusual magnetic field in the spectra.

Another possibility, recently explored by \citet{Fuller2021}, is that in stars with self-excited pulsations, tidal interactions with unstable modes can drive stars away from synchronicity rather than toward it, leading to very low (or in other cases, very high) rotation rates. This process is predicted to be most efficient in binaries with periods of a few days and is thus a potentially promising avenue for explaining the slow rotation of the B star in NGC 2004\,\#115. 

It is somewhat curious that among NGC 2004\,\#115 and the three non-synchronized objects listed above with similar periods, 3 of the 4 systems are triples. Given that a tertiary provides a natural mechanism to perturb the inner orbit, it is worth considering whether there are conditions under which tidal dynamics in triples can support non-synchronous rotation. On the other hand, a large majority of all close binaries have outer tertiaries \citep[e.g.][]{Tokovinin2006}, so the two issues may be unrelated.

\subsection{The search for dormant black holes BHs}

NGC 2004\,\#115 is not the first, and probably will not be the last, dormant BH imposter with an inferred inclination that is close to face-on. When considering such candidates, two facts should be taken into account:

\begin{enumerate}
    \item {\it Dormant BH binaries are an intrinsically rare population.}  This must be accounted for when comparing models that include a BH to alternative models which appear improbable on some other grounds. For a typical binary in a 3 day orbit, tidal synchronization is a reasonably safe assumption: it is empirically true that most stars in binaries with periods this short are tidally synchronized. It is a very risky assumption, however, on which to base an argument for a rare scenario such as a dormant 25\,$M_{\odot}$ BH, particularly one in an improbable configuration as proposed in NGC 2004 \#115. Such an argument (roughly) implies that 25\,$M_{\odot}$ BH companions in close binaries are more common than non-synchronized close binaries. 
    
    A similar argument applies to the assumption that the B stars in LB-1, HR 6819, and NGC 1850 BH1 were main-sequence stars with masses $M\gtrsim 5\,M_{\odot}$. It is empirically true that a large majority of stars with temperatures and radii similar to the observed B stars in these systems are indeed main-sequence stars with $M\gtrsim 5\,M_{\odot}$. But it is probably {\it not} true that close, dormant BH companions to B stars are more common than stripped B stars, which have similar temperatures and radii but much lower masses. In short: priors!

    \item {\it Low inclinations both are intrinsically rare and can make any single-lined binary a BH candidate}. The inference of a large unseen companion mass from the assumption of tidal synchronization is a familiar issue. We are reminded, for example, of the search for massive companions to hot subdwarfs in close binaries by \citet{Geier2010}, who used $v \sin i$ measurement and the assumption of tidal synchronization to infer inclinations and companion masses. They found 6 subdwarfs with inferred companion masses that suggested BH or neutron star companions. The inferred inclinations of these objects were all very low: 16, 14, 26, 14, 23, and 27 deg. The probability of all high-mass companions just happening to have such low inclinations is less than 1 in a million, so it is much more likely that some systems are just not synchronized, and in these cases one will infer a high companion mass if one assumes synchronization. 

 Any single-lined binary becomes a BH candidate if the inferred inclination is low enough. That is, it is easy to get a large number when dividing by 0. But low inclinations are geometrically disfavored: only 13\% of binaries should have inclinations below 30 degrees, and binaries selected from photometric or RV surveys tend to be biased toward {\it high} inclination.  We thus advocate extreme scepticism of dormant BH candidates with low inferred inclinations, particularly if the data would be consistent with a normal stellar companion for a higher inclination. 
 
Binary stellar evolution models provide another useful check on the fidelity of BH candidates \citep[e.g.][]{Stevance2021}. BH formation models are subject to significant uncertainties, so a lack of theoretical models for an observed candidate does not in itself make the candidate untenable. However, binary evolution models can be particularly useful for identifying non-BH scenarios that can match the observed properties of a BH candidate, as was possible in NGC 1850 BH1, LB-1, and HR 6819.
\end{enumerate}

\section*{Acknowledgements}
We thank the referee for a constructive report. We thank Danny Lennon, Tomer Shenar, Jim Fuller and Antonio Rodriguez for helpful discussions. 

\section*{Data Availability}
Data used in this study are available upon request from the corresponding author. 



\bibliographystyle{mnras}

\appendix
\section{Radial velocities}
\label{sec:rvdetails}

\begin{table}
\caption{RVs of the B star}
\label{tab:rvs_s_B}
\begin{tabular}{lcc}
JD $-$ 2400000.5 & ${\rm RV}_{{\rm this\,work}}$  & ${\rm RV}_{{\rm this\,work}}-{\rm RV}_{{\rm Lennon+21}}$ \\
\hline
 & [$\rm km\,s^{-1}$] &  [$\rm km\,s^{-1}$] \\
\hline
52982.20114 & $281.3 \pm 2 $ & 8.7 \\
52982.22809 & $284.1 \pm 2 $ & 1.7 \\
52982.25504 & $288.0 \pm 2 $ & 3.0 \\
52988.21908 & $307.4 \pm 2 $ & 13.4 \\
52988.25059 & $311.3 \pm 2 $ & 19.0 \\
52988.27754 & $315.9 \pm 2 $ & 19.9 \\
53005.13673 & $264.3 \pm 1 $ & 1.3 \\
53005.16375 & $265.5 \pm 1 $ & -0.0 \\
53005.19070 & $264.2 \pm 1 $ & -2.7 \\
53005.22133 & $268.3 \pm 1 $ & -2.7 \\
53006.13530 & $369.9 \pm 1 $ & 0.2 \\
53006.16225 & $372.8 \pm 1 $ & -4.0 \\
53006.18921 & $375.0 \pm 1 $ & -2.3 \\
53008.04798 & $265.2 \pm 1 $ & -3.1 \\
53008.07501 & $265.8 \pm 1 $ & 1.3 \\
53008.10197 & $266.6 \pm 1 $ & -1.0 \\
53008.16305 & $269.4 \pm 1 $ & -0.3 \\
53008.19000 & $271.7 \pm 1 $ & 0.0 \\
53008.21695 & $272.4 \pm 1 $ & -3.0 \\
53009.16182 & $379.3 \pm 1 $ & 2.0 \\
53009.18877 & $380.6 \pm 1 $ & -1.3 \\
53009.21573 & $383.5 \pm 1 $ & 1.1 \\
53012.09557 & $381.9 \pm 1 $ & -0.2 \\
53012.12259 & $384.0 \pm 1 $ & 0.6 \\
53012.14954 & $385.3 \pm 1 $ & 2.2 \\
53012.18344 & $387.9 \pm 1 $ & 1.9 \\
53012.21045 & $388.4 \pm 1 $ & 2.6 \\
53012.23274 & $390.3 \pm 1 $ & 2.5 \\
52955.26626 & $242.5 \pm 1 $ & -4.1 \\
52955.29327 & $240.4 \pm 1 $ & -4.2 \\
52955.32022 & $239.1 \pm 1 $ & -3.0 \\
52989.25927 & $371.2 \pm 1 $ & -0.7 \\
52989.28623 & $369.2 \pm 1 $ & -3.9 \\
52989.31318 & $365.8 \pm 1 $ & -5.7 \\
\hline
\end{tabular}
\end{table}

\begin{table}
\caption{RVs of the Be star}
\label{tab:rvs_Be}
\begin{tabular}{lc}
JD $-$ 2400000.5 & ${\rm RV}_{{\rm this\,work}}$ \\
\hline
 & [$\rm km\,s^{-1}$] \\
\hline
52982.23 & $284 \pm 15 $ \\
52988.25 & $284 \pm 9 $ \\
53005.18 & $288 \pm 6 $ \\
53006.16 & $305 \pm 11 $ \\
53008.07 & $304 \pm 9 $ \\
53008.19 & $282 \pm 24 $ \\
53009.18 & $311 \pm 25 $ \\
53012.17 & $287 \pm 34 $ \\
52955.29 & $326 \pm 18 $ \\
52989.29 & $298 \pm 13 $ \\
\hline
\end{tabular}
\end{table}

Table~\ref{tab:rvs_s_B} lists our measured RVs for the B star, and Table~\ref{tab:rvs_Be} lists values for the Be star. 

The formal RV fitting uncertainties for the B star are typically very small ($\lesssim 0.1\,\rm km\,s^{-1}$). However, we found that we often measured RVs for different regions of the same spectrum that differed by up to a few $\rm km\,s^{-1}$, larger than the formal uncertainties. To investigate potential systematics in the wavelength solution that could explain this, we also fit the RVs of another narrow-lined star observed by the VLT/FLAMES survey of NGC 2004, NGC 2004 \#86 \citep{Evans2006}. This star does not appear to be a binary, but we still found small shifts in our measured RVs across epochs and wavelength ranges, with an RMS variability of $0.9\,\rm km\,s^{-1}$. We therefore conservatively adopt a minimum uncertainty of $1\,\rm km\,s^{-1}$ for the RVs of the B star in NGC 2004 \#115. For observations with the HR 02 grating, we find somewhat larger scatter in the measured RVs of both  NGC 2004 \#86 and the B star in NGC 2004 \#115, so we adopt an uncertainty of $2\,\rm km\,s^{-1}$.

RV uncertainties for the Be star are larger, and its RVs are not found or expected to change significantly across individual exposures within an observing block. We therefore fit individual exposures of each observing block separately, and report the mean and RMS value across these exposures as the star's measured RV and uncertainty. 

\begin{figure}
    \centering
    \includegraphics[width=\columnwidth]{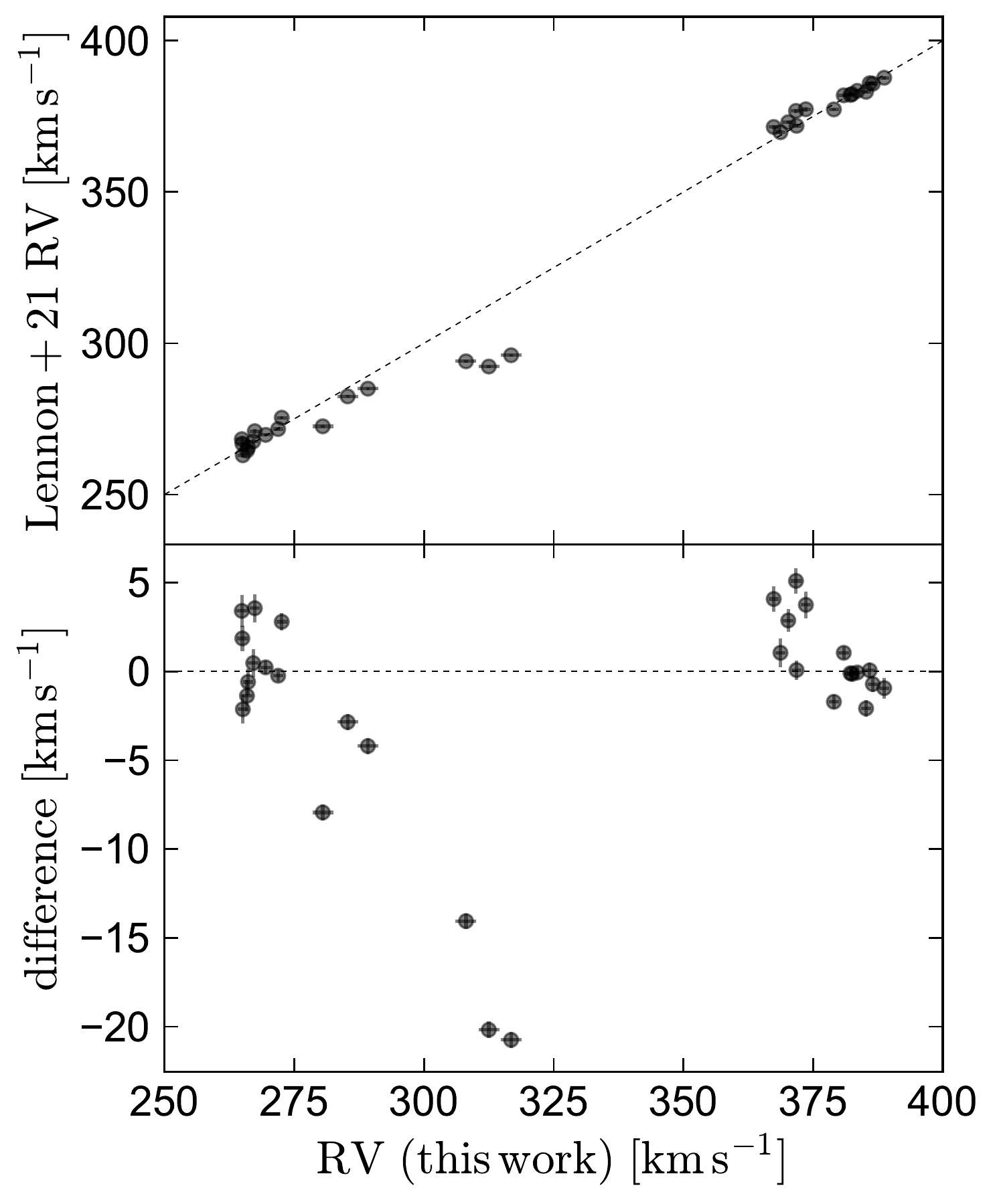}
    \caption{Comparison of our RV measurements for the narrow-lined star and those reported by \citet{Lennon2021}. In several epochs, there are highly significant differences. In those cases, the RV reported by \citet{Lennon2021} is inconsistent with the data (see Figure~\ref{fig:rvfit_compariosn}).  }
    \label{fig:rvs_difference}
\end{figure}

\begin{figure*}
    \centering
    \includegraphics[width=\textwidth]{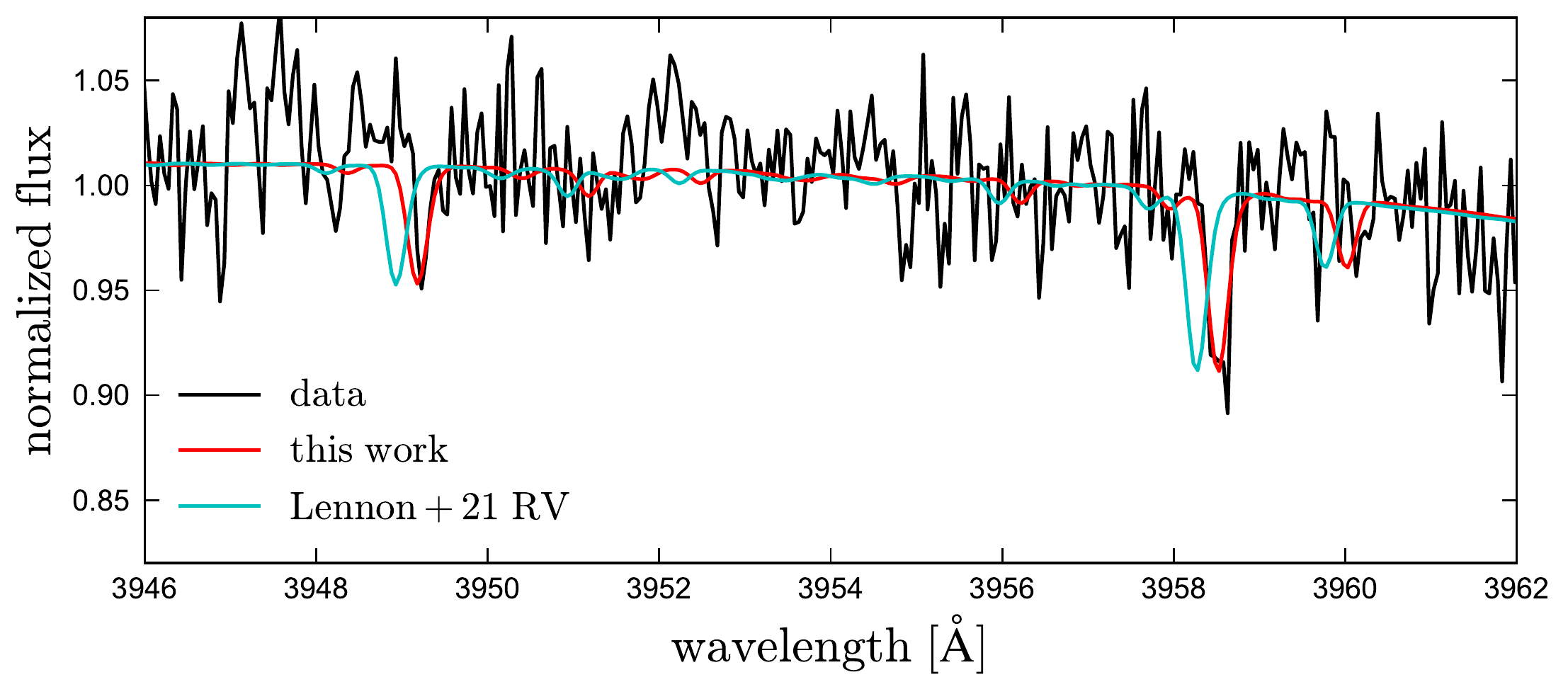}
    \caption{Comparison of our measured RV and the value reported by \citet{Lennon2021} in HR 02 exposure 6, one of the epochs in which the RVs are most discrepant. The RV reported by \citet{Lennon2021} misses the narrow metal lines.  }
    \label{fig:rvfit_compariosn}
\end{figure*}

Our measured RVs are compared to those from \citet{Lennon2021} in Figure~\ref{fig:rvs_difference}, and Figure~\ref{fig:rvfit_compariosn} compares the model spectra and data for one particularly discrepant case.

\begin{figure*}
    \centering
    \includegraphics[width=\textwidth]{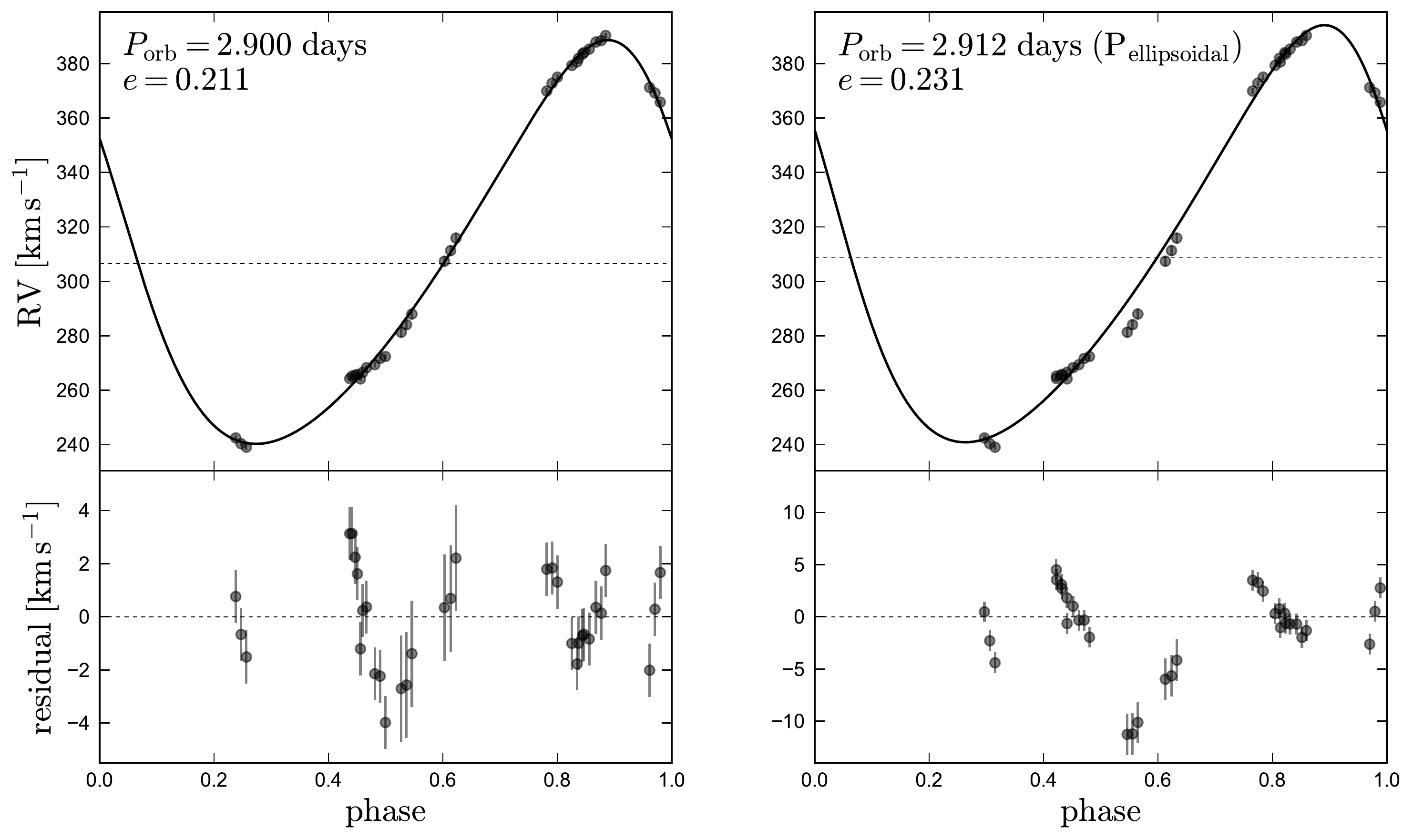}
    \caption{Binary (as opposed to triple) fits to the RVs of the B star. In the left panel, we leave the orbital period free. We obtain a reasonably good fit, but there is still structure in the residuals. In the right panel, we fix the orbital period to the photometric value. Here the fit is not satisfactory. This leads us to consider triple solutions. }
    \label{fig:rvs_comparisons}
\end{figure*}

Figure~\ref{fig:rvs_comparisons} shows best-fit pure binary solutions to the RVs of the B star when the period is left free (left) or fixed to the photometric period (right). Parameters of these solutions, as well as those considered by \citet{Lennon2021}, are listed in Table~\ref{tab:RVs_summary}.

\begin{table*}
\caption{Comparison of RV solutions. The triple model (Figure~\ref{fig:rvs}) fits the B star's RVs with a hierarchical triple solution. ``Binary (free $P_{\rm orb}$)'' fits the same RVs with a binary solution and allows the period to vary, while ``Binary (fixed $P_{\rm orb} = P_{\rm ellipsoidal}$)'' fixes the orbital period to the period of ellipsoidal variation measured from the light curve. ``Binary (Lennon+21, subset of VLT epochs)'' is the fiducial solution from \citet{Lennon2021}, which only fits the RVs obtained within a 1-week period within which the effects of the tertiary are negligible. ``Binary (Lennon+21, all VLT epochs)'' is the binary solution obtained by \citet{Lennon2021} when all the epochs were included in the fit. } 
\label{tab:RVs_summary}
\begin{tabular}{lllll}
Model & $P_{\rm orb,\,inner}\,\,[{\rm day}]$   & $e$ & $K_{B,\,\rm inner}\,\,[\rm km\,s^{-1}]$ & rms\,$[\rm km\,s^{-1}]$ \\
\hline
Triple (fiducial)  &  2.912  &  0 & 62.6  & 0.8 \\
Binary (free $P_{\rm orb}$)  &  2.900  &  0.21  & 74.1 & 1.8     \\
Binary (fixed $P_{\rm orb} = P_{\rm ellipsoidal}$)  &  2.912  &  0.23 & 76.6   & 4.1       \\
Binary (Lennon+21, subset of VLT epochs)  &  2.918  &  0  & 62.4 & 1.2   \\
Binary (Lennon+21, all VLT epochs)  &  2.88  &  0.27  & 74 & 3.5   \\

\hline
\end{tabular}
\end{table*}

\label{sec:rv_comparison}

\bsp	
\label{lastpage}
\end{document}